# Observation of light nuclei formation as nuclear coalescence in CC-interactions at 4.2 AGeV/c

**K. H. Khan**[1]
*COMSATS Institute of Information Technology, Islamabad, Pakistan*
E-mail: `kamal.khan@cern.ch`

**M. K. Suleymanov**
*\*) COMSATS Institute of Information Technology, Islamabad, Pakistan*
*\*\*)Veksler and Baldin Laboratory of High Energy Physics, JINR ,Dubna 141980, Russia*
E-mail: `mais_suleymanov@comsats.edu.pk`

**M. Ajaz**
*\*) COMSATS Institute of Information Technology, Islamabad, Pakistan*
*\*\*)Abdul Wali Khan University Mardan, Pakistan*
E-mail: `Muhammad.Ajaz@cern.ch`

**Ali Zaman**
*COMSATS Institute of Information Technology,Islamabad, Pakistan*
E-mail: `ali_zaman@comsats.edu.pk`

**Sh. Khalilova**
*Institute of Physics NAS,Baku, Azerbaijan*
E-mail: `shahla.ganbarova@cern.ch`

The average multiplicity of light nuclei and $\pi^-$-mesons, emitted in $He^{12}C$- and $C^{12}C$-interactions at *4.2 A GeV*/c were studied as a function of number of identified protons. In both interactions, the behaviour of average multiplicity of $\pi^-$-mesons are in agreement with results coming from the Cascade model. The model could not describe the behaviour of average multiplicity of light nuclei produced in $He^{12}C$-interactions. In case of $C^{12}C$ –interactions the model could describe qualitatively the behaviour of the average multiplicity of light nuclei. An essential deviation was observed in some of the most central events. We believe that nuclear coalescence effect may be a reason of this deviation.

# 1. Introduction

In nuclear- nuclear interactions, light nuclei emitted at rapidities close to that of beam or target are fragments of colliding nuclei. Light nuclei and antinuclei, are, however, being emitted in central kinematic region, at central rapidities. The dominant mechanism may be nuclear coalescence effect. Nucleons which have found themselves very close in phase space may form nuclei and the process is known as nuclear coalescence effect [1].

The coalescence model was initially developed for deuteron production considering phase-space probability distributions of the proton and neutron [2]. Then the phase-space relation was extended for heavier clusters of mass A [3] and now more general form of phase space relation [4] is

$$E_A \frac{d^3 n_A}{dp_A^3} = B_A (E_p \frac{d^3 n_p}{dp_p^3})^A, \qquad (1)$$

Where $B_A$ is coalescence parameter and is

$$B_A = (\frac{2S_A+1}{2^A}) \frac{1}{N!} \frac{1}{Z!} (R_{np})^N (\frac{4\pi}{3} P_0^3)^{A-1}$$

$S_A$ is the spin of the cluster of mass A, and N and Z is the neutron and proton numbers of the composite particle. The factor $R_{np}$ is the ratio of neutrons to protons participating in the collision and $P_0$ is momentum radius within which nucleon pairs will fuse.

Some other approaches were studied where coalescence parameter $B_A$ has strong dependence on volume of the interaction zone [5].The relation in equation (1) has been verified at Bevalac and AGS energies. The coalescence paramètre $B_A$ has shown an increase with transverse cluster mass $M_t$, and has no dependence on centrality [6]. Variation of $B_A$ with system, centrality, and collision energy were discussed [7]. Their results are inconsistent with model they used, and could not be described the coalescence effect. The results of measurements of light nuclei from *A*=1 to *A*=7 have shown in [8]. They are unable to extract information about the coalescence effect. Different mechanism for light nuclei formation in different kinematical regions, Fragmentation at low rapidities and coalescence at mid rapidities was observed [9]. The different aspects of coalescence parameter have studied, and no clear information about the coalescence affect was obtained [10, 11,12].

These results demonstrate that light nuclei production, in nuclear collisions, is an important effect which can give some essential information about the formation of complex baryon system (light nuclei) and to understand the states of matter just after the Big Bang. Mainly, multi nucleon interactions caused light nuclei formations at normal nuclear temperature, however, nuclei could be formed as a result of nuclear coalescence effect under high pressure but at normal temperature after the Big Bang. In nuclear collisions, light nuclei, emitted as a result of fragmentation must decrease with centrality and should have minimum at the most central event and it is possible that light nuclei may be formed as a result of coalescence effect and will appear in central events. So the study of centrality dependences of light nuclei formation in relativistic light nuclear collisions can give clear information about the coalescence mechanism of nuclei formation. Light nuclei interactions were considered to get simple physical picture as compared heavy ion collisions.

In this paper the light nuclei emission was studied as a function of collision centrality in He$^{12}$C- and C$^{12}$C - interactions at 4.2AGev/c. Study of some signatures of nuclear coalescence effect is the main target of investigation. The Experimental data was obtained from 2-m propane bubble chamber of LHE, JINR and compared with simulated data coming from the Cascade model.

**2. Methodology**

The experimental data on He$^{12}$C and $^{12}$C$^{12}$C interactions at 4.2 AGev/c were obtained from 2m propane bubble chamber, JINR, Russia .The chamber was placed in a 1.5 T magnetic field, and irradiated with the beams of light relativistic nuclei at the Synchrophasotron. Practically all secondary particles emitted at $4\pi$ total solid angle were detected in the chamber. The $\pi^-$ -mesons were identified quite well in the propane chamber. The average minimum momentum for pion registration was set to about 70 MeV/c. The protons were selected by the statistical method applied to all positive particles with momentum of p >150 MeV/c. The experiment could measure only deutron and some mixture of light nuclei (for details see [13]).
The average multiplicity of light nuclear fragments was studied as a function of centrality, and centrality was fixed by the number of identified protons in an event. The experimental results were compared with results coming from simulation with Cascade model.

Cascade model [14] is the most popular model which is used to describe the general features of relativistic nucleus-nucleus collisions. It is an approach based on simulation using Monte-Carlo techniques and is applied to situation where multiple scattering is important. The cascade model does not include any medium or collective properties and each colliding nuclei is treated as a gas of nucleon bound in a potential well. The Pauli principle and the energy momentum conservation are obeyed in each inter-nuclear interaction. The remaining excited nuclei, after the cascade stage are described by the statistical theory in the evaporation approximation. Model includes nuclear fragments from projectile and target, but does not include nuclear coalescence effect [ 15, 16, 17].

In the experiment, 39544 events of $C^{12}C_3H_8$, and 22975 events of $He^{12}C_3H_8$ interactions were used at 4.2 A GeV/c. In the case of cascade code 40000 events of both $C^{12}C$ and $He^{12}C$ interaction were used.

## 3. Results and discussions

The average multiplicity of the negative pions ($<\pi^->$) in $He^{12}C$- and $C^{12}C$-interactions at 4.2 A GeV/c was shown in figure1 (a and b), as a function of the number of identified protons ($N_p$) in an event (centrality). The open circles represent experimental data and the corresponding solid circles represent the simulated data. The average multiplicity of $\pi^-$-mesons$<\pi^->$ is increasing with centrality except at the most central events, where a small decrease was shown. The cascade code is in good agreement with experiment for pion production.

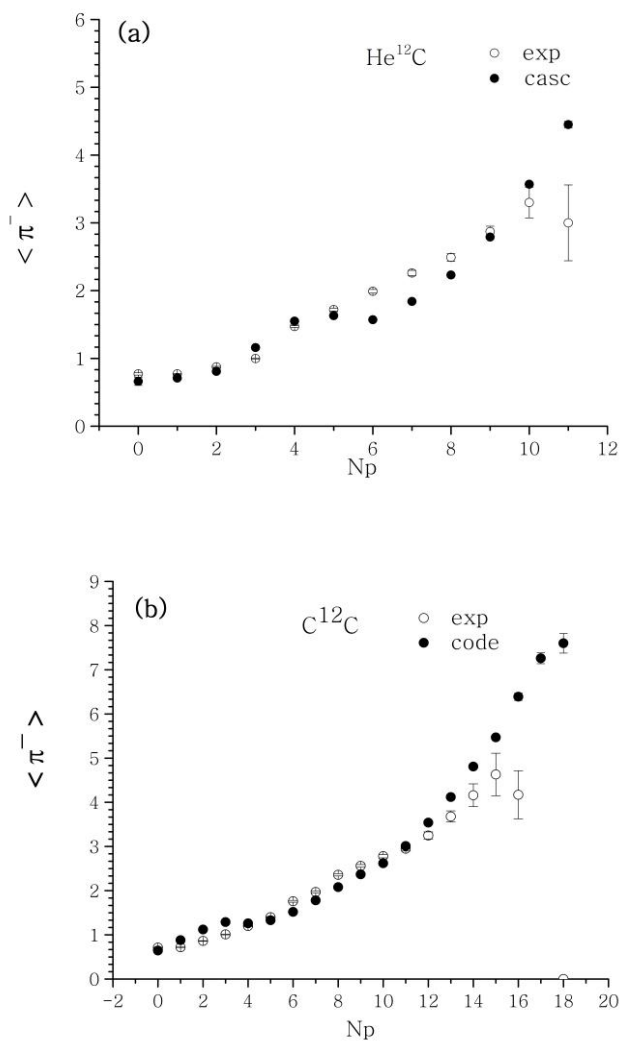

*Fig.1. Average multiplicity of $\pi$-mesons as number of proton (a) $He^{12}C$-inteactions (b) $C^{12}C$-interactions.*

The values of the average multiplicity of the light nuclei (<N>), produced in $He^{12}C$- and $C^{12}C$-interactions at 4.2 A GeV/c as a function of the $N_p$ were shown in figure 2(a and b). The open circles indicate experimental data, whereas the solid circles represent simulated data. The simulated data was normalized to first point for comparison. One can see that the model could not describe the behavior of the <N> as a function of $N_p$ for the $He^{12}C$-interactins. Some part of the difference may be, due to methodical problems of the experiment. In experiment light nuclei could be measured as some mixture of deuterons, tritium and other fragments, whereas cascade model can separate them completely. In figure 2(b) the experimental results have similar behavior as model except in the most central region where a small increase was observed strangely. The strangeness is, with centrality multiplicity of light nuclei <N> must decrease or saturates. It cannot increase at maximum centrality as it follows from model. We think the reason of increase in the multiplicity of light nuclei at central events is additional formation of light nuclei in dense nuclear medium. In this medium protons and neutrons which are very close in phase space may form light nuclei and this process is known as nuclear coalescence effect.

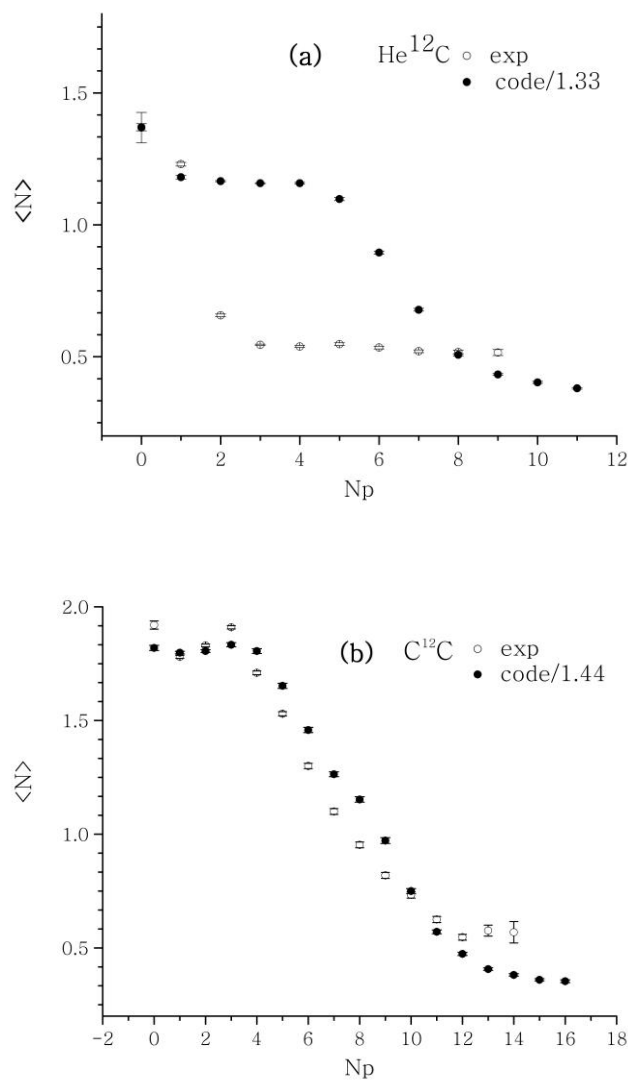

*Fig.2. Average multiplicity of light nuclei as number of proton (a) He$^{12}$C-interactions (b) C$^{12}$C-interactions.*

It was observed from figure 1(b) and figure 2(b) that deviation in multiplicities of $\pi^-$-mesons and light nuclei is in same region systematically, which may confirm some effect not methodical problems, and the effect may be the nuclear coalescence effect.

**4. summary**

The average multiplicity of light nuclei and $\pi^-$ -meson emitted in *He$^{12}$C-* and *C$^{12}$C* -interactions at *4.2 A GeV*/c as a function of centrality were studied. The centrality of collisions was fixed by the number of identified protons. The behaviors of the average multiplicity of $\pi^-$ -meson as a function of number of protons are in agreement with results coming from the Dubna version of cascade model in both interactions. The model could not describe qualitatively and quantitatively the behaviors of the average multiplicity for light nuclei produced in He$^{12}$C-interactions as a function of centrality. In case of C$^{12}$C – interaction the model could describe qualitatively the behaviors of the average multiplicity for light nuclei. An essential deviation was observed only in most central events when the number of identified protons is limited high – great than 12. In this area data coming from the model decrease and saturates. But for the experimental data a small increase was observed. The result is strange because the sharp decrease was being expected for the light nuclear multiplicity in this area. We believe that nuclear coalescence effect may be a reason of this increase. Although some methodical problems connected with light nuclei identification may give some contribution in this region. To get additional information and confirm the results we are going to do more methodical work and include some other simulation code.

**5. References**

[1]  Introduction to Relativistic Heavy Ion Physics, J.Bartke, world scientific..

[2] S.T Butler and C.A. Pearson, Phys.Rev.Lett. **7**,69(1961);Phys.Rev **129**,836 (1963).

[3] A. Schwarzschild and C. Zupancic, Phys. Rev. **129**, 854 (1963).

[4] S. Albergo et al. Phys. Rev. C, v. **65**, 034907

[5] J. l Nagle et al. Phys. Rev.C **53**(1996).

[6] I.G. Bearden et al. European Physical Journal C ; EPJC011124.

[7] M.J. Bennet et al., Phys. Rev.C **58** (1998).

[8] T. A. Armstrong et al. Phys. Rev.C **61**, 064908(2000).

[9] S.S. Adler et al. Phys. Rev. Lett. **94,** 122302(2005).

[10] H.Aidong Liu, J. Phys. G: Nucl. Part. Phys. **34** (2007) S1087–S1091

[11] J. Barrette et al., Phys. Rev. Lett. **73**,2532 (1994). E802 Collaboration.

[12] Zhangbu Xu for the E864 Collaboration. arXiv: nucl-ex/9909012.

[13] N.Akhababian et al.,JINR Preprint 1-12114,1979.

[14] M.I. Adamovich, et al., Z. Phys, **A 458** (1997) 337.

[15] V.S Barashenkov, F.Zh. Zheregi, Zh.Zh. Muslmanbekov, JINR preprint, P2-83-117 Dubna, 1983.

[16] V.S Barashenkov and V.D Toneev, "interaction of High energy particles and atomic nuclei with nuclei" , Moscow ,Atomizadt, 1972.

[17] V.D Toneev ,. Gudima. K. K. Nucl. Phys. **A400**,(1983) 173.